\documentclass[final,3p,times,twocolumn,sort&compress]{elsarticle} 
\usepackage{graphicx}
\usepackage{amssymb}
\usepackage{lineno}
\usepackage{amsmath}
\usepackage{tikz}
\usepackage{hyperref}
\usetikzlibrary{shapes.geometric, arrows}
\newcounter{bla}

\journal{Computer Physics Communications}

\begin{document}
\begin{frontmatter}
\title{GPU-Accelerated MATLAB Software for Atom-Ion Dynamics}
\author[a]{Saajid Chowdhury\corref{author}}
\author[a]{Jes\'us P\'erez-R\'ios}
\cortext[author] {Corresponding author.\\\textit{E-mail address:} saajid.chowdhury@stonybrook.edu}
\address[a]{Department of Physics and Astronomy, Stony Brook University, Stony Brook, New York 11794, USA}
\begin{abstract}
    We present a MATLAB script which can use GPU acceleration to simulate a trapped ion interacting with a low-density cloud of atoms. This script, called \verb+atomiongpu.m+, can massively parallelize MD simulations of trajectories of a trapped ion and an atom starting far away. The script uses \verb+ode45gpu+, which is our optimized and specialized implementation of the Runge-Kutta algorithm used in MATLAB's ODE solver \verb+ode45+. We first discuss the physical system and show how \verb+ode45gpu+ can solve it up to 22x faster than MATLAB's \verb+ode45+. Then, we show how to easily modify the inputs to \verb+atomiongpu.m+ to account for different kinds of atoms, ions, atom-ion interactions, trap potentials, simulation parameters, initial conditions, and computational hardware, so that \verb+atomiongpu.m+ automatically finds the probability of complex formation, the distribution of observables such as the scattering angle and complex lifetime, and plots of specific trajectories. 
\\

\noindent {\bf PROGRAM SUMMARY}

\begin{small}
	\noindent
	{\em Program Title:} \verb+atomiongpu.m+\\
	{\em CPC Library link to program files:} \\ 
    {\em Developer's repository link:} \href{https://github.com/saajidchowdhury/supplementGPU}{https://github.com/saajidchowdhury/supplementGPU}\\
	{\em Licensing provisions:} \\ 
	{\em Programming language:} MATLAB R2023a, with Parallel Computing Toolbox installed\\
    {\em Supplementary material:} \href{https://github.com/saajidchowdhury/supplementGPU}{https://github.com/saajidchowdhury/supplementGPU}\\
    {\em Nature of problem:} Simulate classical dynamics (Newton's laws) of an ion and atom, with up to several million different sets of initial conditions, store the final conditions and a few other scalar observables and their distributions, and plot specific trajectories. \\ 
	{\em Solution method:} Implementing the algorithm behind \verb+ode45+, MATLAB's fourth/fifth-order adaptive-timestep Runge-Kutta method for propagating ordinary differential equations, we write a single, self-contained function, \verb+ode45gpu+. Then, we use MATLAB's \verb+arrayfun+ to parallelize it on multiple CPUs or GPUs. Finally, we wrote the script \verb+atomiongpu.m+ for quickly and conveniently using \verb+ode45gpu+. \\ 
    {\em Additional comments:} The source code for \verb+atomiongpu.m+, \verb+ode45gpu+, and figures can be found on the repository.\\ 
	\\
\end{small}
\end{abstract}
\end{frontmatter}
\begin{keyword}
	MATLAB; GPU; Molecular dynamics; Atom-ion interactions; \verb+ode45+
\end{keyword}

\section{Introduction}

In many fields such as chemical physics, biophysics, and materials science, molecular dynamics (MD) is a commonly used technique to make predictions. Typically, MD simulates the motion of the system's atoms and molecules, following Newton's laws, for some amount of time. The results of MD simulations reveal properties of the system, such as thermal distributions and reaction rates \cite{Pinkas2020,Hirzler2020}.

In atomic and molecular physics, one specific application of MD is the interaction of a trapped ion and free atom, essential for cold and ultracold chemistry \cite{Tomza2019}. In experimental studies of atom-ion interactions, Paul traps have been used to trap ions in rotating quadrupole electric fields, and buffer gases have been used to cool down ions enough to isolate quantum mechanical aspects of atom-ion scattering \cite{Feldker2020,Walewski2025,Pinkas2023}. In addition, trapped ions have been observed to temporarily weakly bind with atoms, forming quasi-bound states known as complexes. Alongside these experimental developments, MD simulations have been used extensively to understand these results and make theoretical predictions, such as the probability of complex formation and thermalization rates \cite{Cetina2012,Hirzler2023,Trimby2022,Furst2018,Pinkas2024,Pinkas2020,Croft2014,Hirzler2020,Londono2022,Londono2023,Yang2017,Liang2025}.

Since the buffer gas typically has a low density, it is safe to assume that the ion interacts with one atom at a time, so only two particles need to be simulated. However, the precise initial conditions of the atom and ion are not known, so it is not enough to run one MD simulation of the atom-trapped-ion system. Typically, several thousand MD simulations are run, each with different initial conditions, to obtain a comprehensive statistical description. 

Because each atom-ion trajectory is independent, it is possible to parallelize these simulations, so that they finish more quickly. Furthermore, since each trajectory contains only two particles, and the full trajectory usually doesn't need to be stored, very little memory is needed. Thus, the atom-trapped-ion system is an ideal candidate for massively parallelizing MD simulations on a GPU. A GPU contains dozens to hundreds of microprocessors working simultaneously, each capable of doing basic computations with low memory. GPU-accelerated MD is already common \cite{NVIDIA2017,Lee2018,Hayes2021,Berendsen1995,Valiev2010,Colavecchia2014}, but the available software does not offer trapping options and is not adequate for problems in cold and ultracold chemistry. In addition, a simple open-source MATLAB implementation of GPU-accelerated classical MD has never been given.

To run an MD simulation, one must numerically solve an ordinary differential equation (ODE). MATLAB has several methods for solving ODEs, including a fourth/fifth-order adaptive-timestep Runge-Kutta method, \verb+ode45+, based on the Dormand-Prince pair \cite{Dormand1980,Shampine1997,MATLAB}. To run calculations on a GPU, one usually redesigns and re-implements the calculation algorithm in the interface known as CUDA \cite{Januszewski2010,Spiechowicz2015}. Fortunately, some MATLAB functions can also be run on a GPU, utilizing MATLAB's ability to internally translate certain primitive operations into CUDA code. This enables us to write a MATLAB function that numerically solves an ODE, \verb+ode45gpu+, which is able to run on both a CPU and a GPU. 

This paper is organized as follows. First, we give a detailed description of the physical system being simulated, culminating in an ODE of the form $\bold{\dot{y}}=f(t,\bold{y})$ and an expression for the force function $f(t,\bold{y})$. Then, we discuss how \verb+ode45+ would typically be used to solve this system on a CPU, and how the performance can be significantly enhanced merely by writing $f(t,\bold{y})$ in an optimal way. Next, we present our in-house MATLAB implementation of fourth/fifth order adaptive-timestep Runge-Kutta called \verb+ode45gpu+, which outperforms \verb+ode45+ for atom-ion interactions on a single CPU core, as well as multiple CPU cores in parallel. We then discuss how we use \verb+ode45gpu+ on a GPU using \verb+gpuArray+s and \verb+arrayfun+, including how we wrote out $f(t,\bold{y})$. We also discuss the performance of \verb+ode45gpu+ on different kinds of GPUs. Finally, we describe the input and output of \verb+atomiongpu.m+, our MATLAB script which automates and simplifies the usage of \verb+ode45gpu+ to study atom-ion interactions. 

\section{The physical system}

\begin{figure*}
    \centering
    \includegraphics[width=\textwidth]{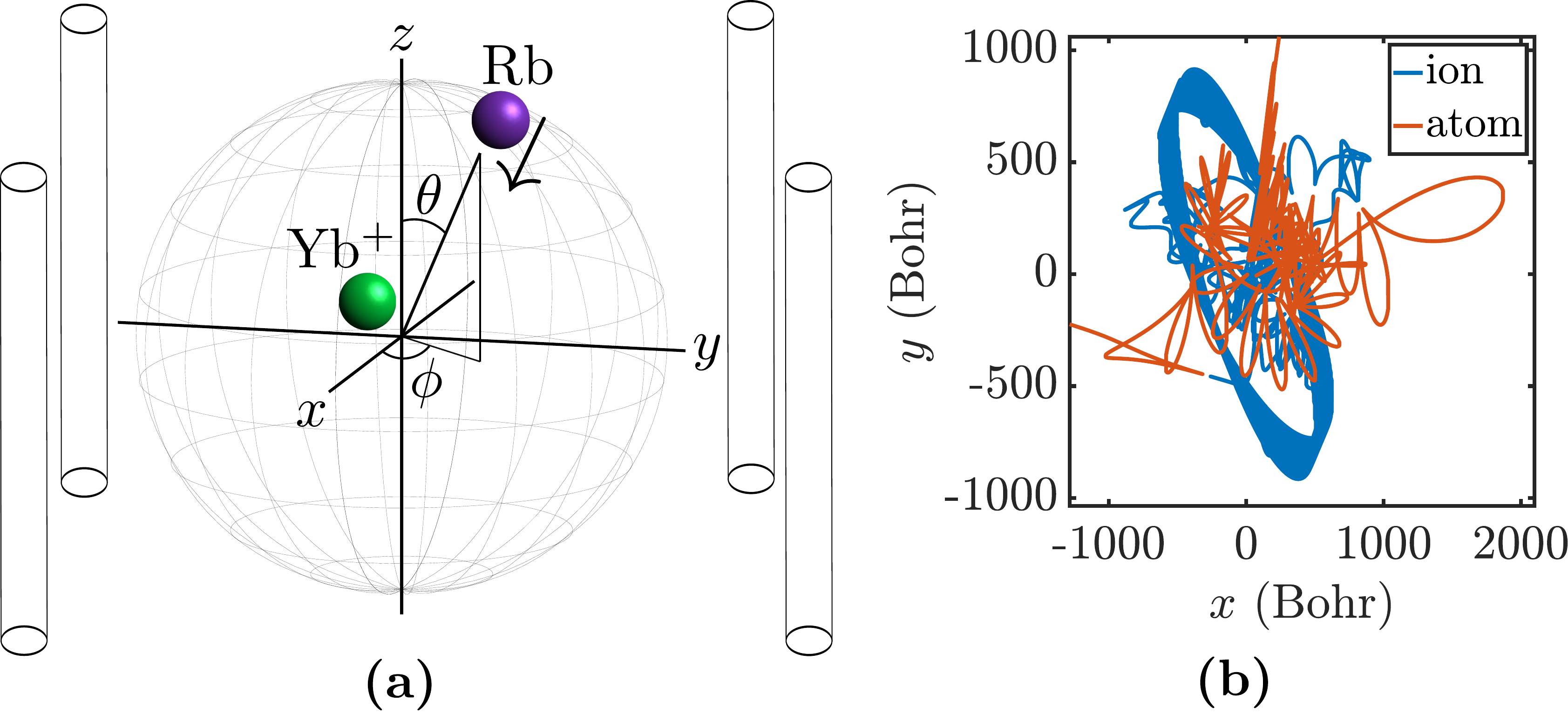}
    \caption{The physical system. Panel (a) shows a schematic of the physical system: an atom launched towards an ion in a quadrupole trap. Panel (b) shows an $xy$-projection of the specially-chosen trajectory used for testing parallelization (the parts of the trajectory where the atom is very far from the ion, in the beginning and end, are not shown). The origin $(0,0,0)$ is the center of the quadrupole trap field.}
    \label{fig1}
\end{figure*}

The MATLAB script \verb+atomiongpu.m+ is able to simulate any system consisting of an atom and ion, which is either untrapped or in a harmonic or Paul trap. It ignores internal degrees of freedom and simulates the system classically, assuming an atom-ion interaction of the form $V(r)=-C_n/r^n+C_m/r^m$. For parallelization, the atom's initial velocity can either be a fixed speed directed towards the ion, or thermally distributed in a random direction; the code can also simulate an individual custom trajectory with customizable initial conditions. However, to benchmark the code, we stick to the following special case. 

\subsection{The benchmark physical system}

The benchmark physical system is a Rb atom, starting at 5,000 Bohr radii ($a_0$) away from a Yb$^+$ ion at rest at the center of a Paul trap. The Rb's initial speed, directed towards the ion, is given by the temperature $T=1\mu$K. This is shown in Fig.~\ref{fig1}(a). The Hamiltonian of the system is 
\begin{align}
    H=\,&\frac{1}{2}m_av_a^2+\frac{1}{2}m_iv_i^2-\frac{C_4}{r_\textrm{ai}^4}+\frac{C_8}{r_\textrm{ai}^8}+\\
        &\sum\limits_{i=1}^{3}(a_i+2q_i\cos\Omega_\textrm{rf}t)\frac{m_\textrm{ion}\Omega_\textrm{rf}^2}{8}r_i^2,
\end{align}
where $m_a$ and $v_a$ are the mass and velocity of the atom, $m_i=m_\textrm{ion}$ and $v_i$ are the mass and velocity of the ion, $r_\textrm{ai}$ is the atom-ion distance, $C_4$ is the charge-induced dipole coefficient and $C_8$ is the short-range coefficient, $a_i,q_i$ for $i=1,2,3$ are trap parameters controlled via the DC potential, distances between metal poles, and the AC potential, and $\Omega_\textrm{rf}$ is the trap radiofrequency used to simultaneously change the polarity of two of the metal poles in the Paul trap. The expression $V_\textrm{ai}(r_\textrm{ai})=-C_4/r_\textrm{ai}^4+C_8/r_\textrm{ai}^8$ is the atom-ion interaction potential. 

Given the Hamiltonian, one can use Newton's laws or Hamilton's equations to find the equations of motion for the atom and ion. These equations can be expressed in the form $\bold{\dot{y}}=f(t,\bold{y})$ where $\bold{y}$ is a 12-dimensional vector, accounting for the 6 degrees of freedom of the system (due to the trap, it is not possible to separate the center of mass from the relative motion). So, $y_{1,2,3}$ are the Cartesian coordinates of the ion, $y_{4,5,6}$ are the ion's velocity components, $y_{7,8,9}$ are the Cartesian coordinates of the atom, and $y_{10,11,12}$ are the atom's velocity components. The force function, $f(t,\bold{y})$, is 
\begin{align}
    &f(t,\bold{y})=(y_4,y_5,y_6,\\
    &              -\frac{1}{m_i}\frac{dV}{dr}\frac{y_1-y_7}{r}-(a_x+2q_x\cos\Omega_\textrm{rf}t)\frac{\Omega_\textrm{rf}^2}{4}y_1,\\
    &              -\frac{1}{m_i}\frac{dV}{dr}\frac{y_2-y_8}{r}-(a_y+2q_y\cos\Omega_\textrm{rf}t)\frac{\Omega_\textrm{rf}^2}{4}y_2,\\
    &              -\frac{1}{m_i}\frac{dV}{dr}\frac{y_3-y_9}{r}-(a_z+2q_z\cos\Omega_\textrm{rf}t)\frac{\Omega_\textrm{rf}^2}{4}y_3,\\
    &              y_{10},y_{11},y_{12},\\
    &              -\frac{1}{m_a}\frac{dV}{dr}\frac{y_7-y_1}{r},\\
    &              -\frac{1}{m_a}\frac{dV}{dr}\frac{y_8-y_2}{r},\\
    &              -\frac{1}{m_a}\frac{dV}{dr}\frac{y_9-y_3}{r}),
\end{align}
where $r=r_\textrm{ai}=\sqrt{(y_1-y_7)^2+(y_2-y_8)^2+(y_3-y_9)^2}$ is the atom-ion distance, and $dV/dr=4C_4/r^5-8C_8/r^9$ is the derivative of the atom-ion potential function. 

The initial conditions are 
\begin{align}
    \bold{y_0}=\,&(0,0,0,0,0,0,\\
                 &r_0\sin\theta\cos\phi,r_0\sin\theta\sin\phi,r_0\cos\theta,\\
                 &-v_0\sin\theta\cos\phi,-v_0\sin\theta\sin\phi,-v_0\cos\theta),
\end{align}
where $r_0=5000a_0$, $v_0=\sqrt{3k_BT/m_a}$, and $\theta,\phi$ are the initial angular coordinates on the sphere of radius $r_0$ centered at the origin. 

Depending on $\theta,\phi$, the trajectory can either be very simple (the atom bounces once and then scatters back away from the ion), or very complicated, such as in Fig.~\ref{fig1}(b), characteristic of the inherent highly non-linear character of the ODE. This complicated trajectory shows that the ion and atom formed a complex, where they are temporarily bound. We specifically chose this trajectory, with $\theta\approx 1.02$ and $\phi\approx 0.22$, to benchmark our code, and sometimes we instead uniformly sampled initial positions on the surface of the first octant of the sphere, with $\theta\in[0,\pi/2]$ and $\phi\in[0,\pi/2]$. By ``uniformly sampled'', we mean that we chose a grid of every combination of 50 evenly-spaced $\phi$-values and 50 $\theta$-values such that each $(\theta,\phi)$ occupies the same solid angle. 

\section{Numerical Implementation}

\subsection{Using ode45 on CPU}

\begin{figure}[h]
    \centering
    \includegraphics[width=0.45\textwidth]{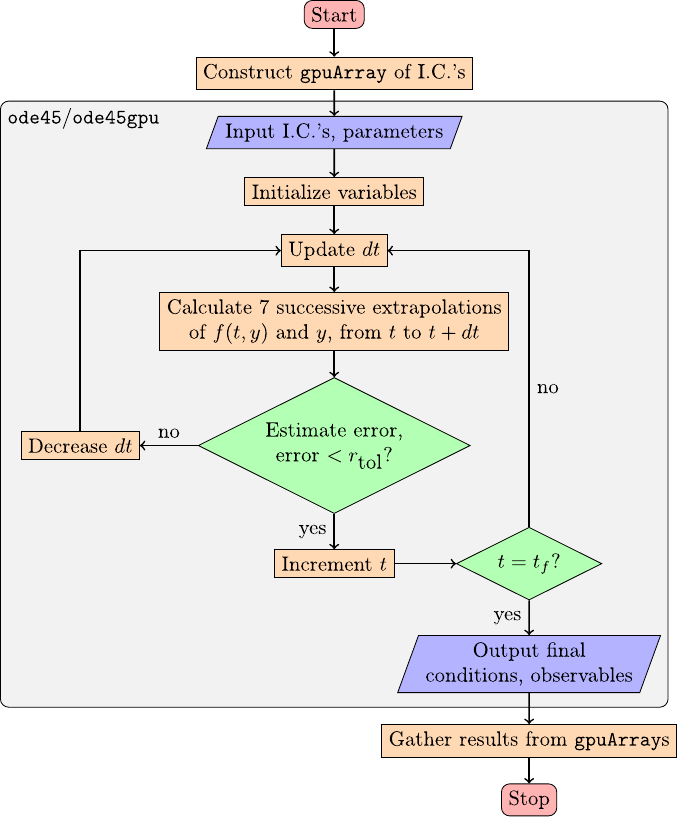} 
    \caption{Flowchart with ode45 and ode45gpu workflow and algorithm. The gray box shows the algorithm for the ODE solver, which is exactly the same for ode45 \cite{Dormand1980,Shampine1997,MATLAB} and ode45gpu, except for the output. Instead of just the final conditions, ode45 outputs the full trajectory, and the observables are computed afterwards. Outside the gray box, in the case of ode45gpu, MATLAB uses a data structure called a gpuArray to store all the initial conditions and results.}
    \label{fig2}
\end{figure}

Usually, to numerically solve $\bold{\dot{y}}=f(t,\bold{y})$, one would use an ODE solver, such as MATLAB's \verb+ode45+. One would provide the solver with the function $f(t,\bold{y})$, the desired simulation time $t_\textrm{max}$, and options such as the absolute and relative tolerances. Then, the solver simulates the integration in discrete steps $dt$ for some finite amount of simulated time, which in our case is $t_\textrm{max}=1$ millisecond. At each timestep, if the estimated error goes above a certain threshold determined by the tolerances, then the adaptive timestep $dt$ is decreased; otherwise, it may stay the same or increase. The layout of this algorithm is shown within the gray box of Fig.~\ref{fig2}. 

\begin{figure}[h]
    \centering
    \includegraphics[width=0.45\textwidth]{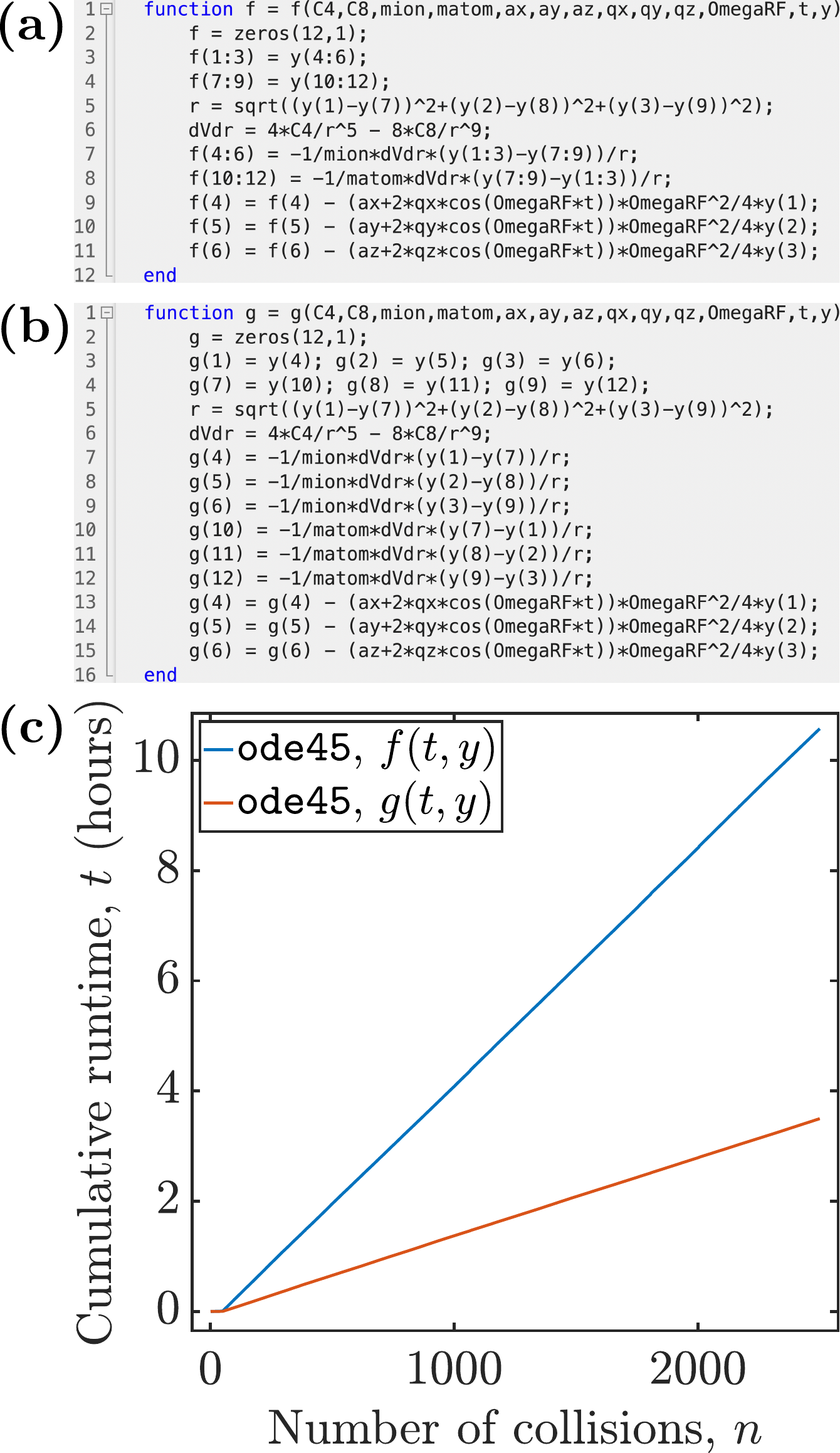}
    \caption{Optimizing the force function. Panels (a) and (b) show two alternatives for the force function. The first one, $f(t,\bold{y})$, is the function given to ode45 on the CPU for the remainder of the paper. The second one, $g(t,\bold{y})$, is an improved version of $f$. Panel (c) shows the runtimes of ode45 with $f$ and $g$ on one CPU core, for 2,500 different trajectories, uniformly sampled on the surface of one octant of a sphere centered at the ion, $\theta,\phi\in[0,\pi/2]$.}
    \label{fig3}
\end{figure}

If one uses the function $f(t,\bold{y})$ in Fig.~\ref{fig3}(a), calculating 2,500 trajectories one at a time on one Intel Haswell 128GB CPU core using \verb+ode45+, the runtime is shown by the blue curve in Fig.~\ref{fig3}(c). It takes \verb+ode45+ over 10 hours to simulate 2,500 different trajectories on one CPU core.

However, if one uses the function $g(t,\bold{y})$ in Fig.~\ref{fig3}(b), then \verb+ode45+ computes exactly the same 2,500 trajectories in only less than 3.5 hours, as shown by the red curve in Fig.~\ref{fig3}(c). So, merely rewriting the force function yields a $\sim 3$x speedup. 

The issue with $f(t,\bold{y})$ is, for example, the appearance of \verb+y(4:6)+ on the RHS of line 3 in Fig.~\ref{fig3}(a). It is wasting time constructing the three-element vector \verb+[y(4),y(5),y(6)]+, only to use each element one by one. Instead of first constructing the vector, it is faster to simply use each element directly, either explicitly written out as in line 3 of Fig.~\ref{fig3}(b), or using a for loop. This explains some of the speedup offered by our function \verb+ode45gpu+.

\subsection{Developing ode45gpu, enhancing CPU runtime}

\begin{figure}
    \centering
    \includegraphics[width=0.45\textwidth]{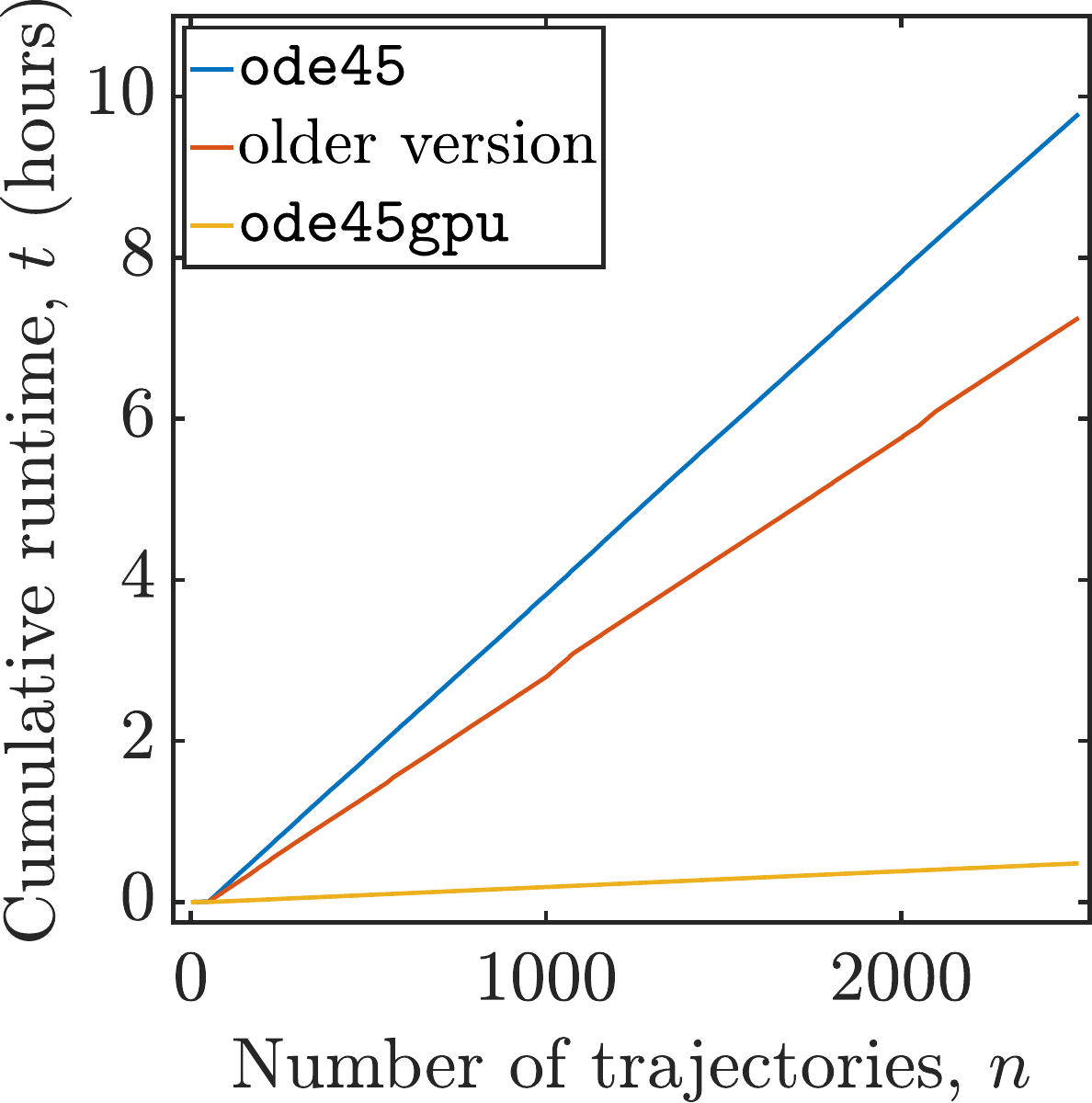}
    \caption{Runtime of ode45 and ode45gpu, as well as a vectorized older version of ode45gpu, running on one CPU, for 2,500 different trajectories, uniformly sampled on the surface of one octant of a sphere centered at the ion, $\theta,\phi\in[0,\pi/2]$.}
    \label{fig4}
\end{figure}


We used the algorithm given in \cite{Shampine1997,Dormand1980} to implement a MATLAB function called \verb+ode45gpu+. This single, self-contained function takes the initial conditions and other parameters as inputs, and outputs only the final conditions (the final values of $t$ and $\textbf{y}$) and other scalar observables. A condition was added to break out of the loop once 1,000,000 timesteps have been computed, to prevent the simulation from taking too long. Since $\textbf{y}$ is a vector with 12 components, an older version of \verb+ode45gpu+ extensively used MATLAB vectors. But, since vectors are not allowed in a function running on the GPU through \verb+arrayfun+, each vector was replaced with 12 variables, leading to the final version of \verb+ode45gpu+. We made sure that \verb+ode45gpu+ and \verb+ode45+ compute the same values for the atom's scattering angle for 2,500 different trajectories. 

\begin{figure}
    \centering
    \includegraphics[width=0.45\textwidth]{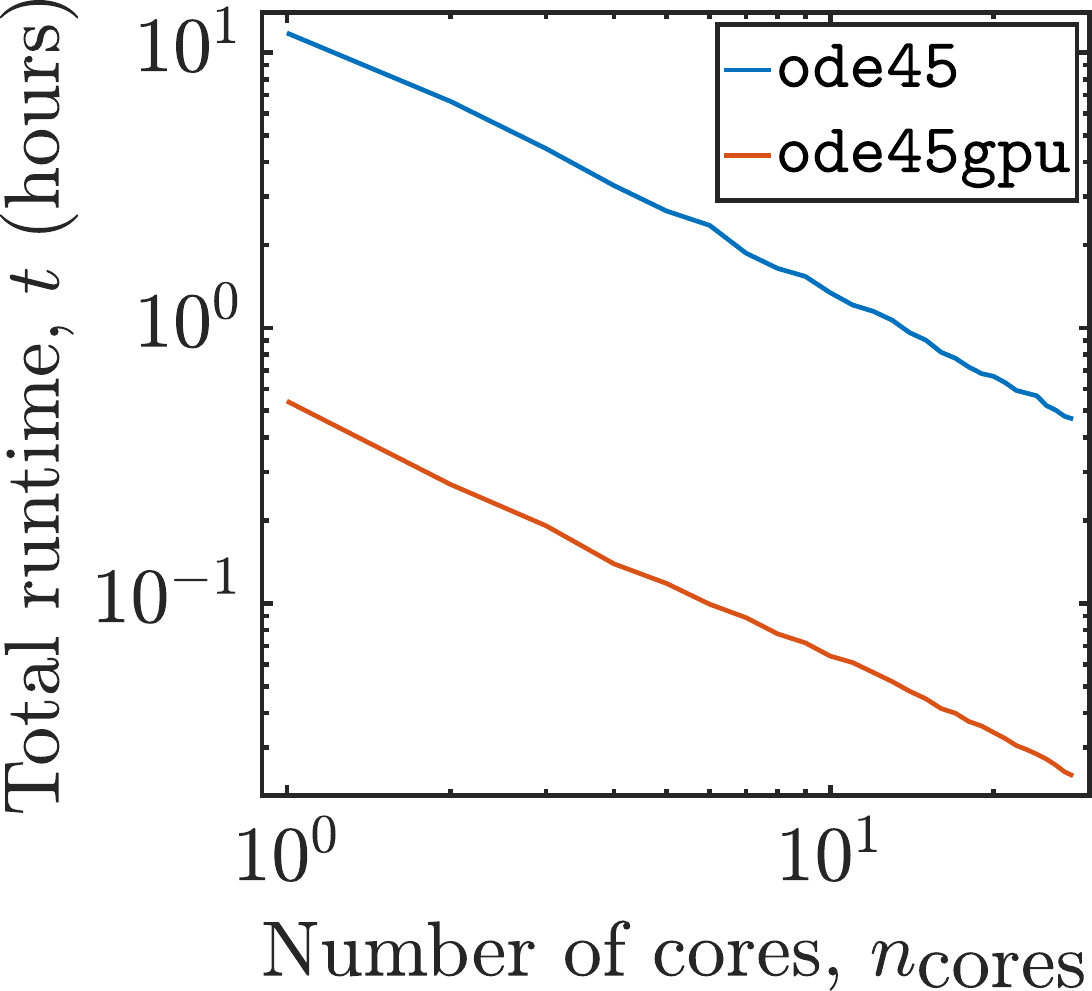}
    \caption{CPU runtime of ode45 and ode45gpu, running a total of 2,500 identical copies of the specially-chosen trajectory shown in Fig.~\ref{fig1}(b), as a function of the number of independent cores, $n_\textrm{cores}$, over which the trajectories are distributed.} 
    \label{fig5}
\end{figure}

Running 2,500 trajectories using \verb+ode45+, \verb+ode45gpu+, and the vectorized ``older version'' of \verb+ode45gpu+ on one CPU core, the runtimes are shown in Fig.~\ref{fig4}. On one CPU core, compared to \verb+ode45+ using $f(t,\textbf{y})$, we find that \verb+ode45gpu+ runs 22 times faster. Furthermore, Fig.~\ref{fig5} shows that when \verb+ode45gpu+ is parallelized across up to 28 cores, this speed ratio is maintained. This stark difference in runtimes is related to the discussion in the previous subsection. 



\subsection{General usage of ode45gpu on a GPU}


Here, we describe how we used \verb+ode45gpu+ on the GPU, giving enough detail so that if a user wants to have more control, they can easily generalize the code for different $f(t,\bold{y})$ and sets of initial conditions. We used MATLAB's \verb+gpuArray+s as input and output, and used \verb+arrayfun+ to execute the function on the GPU. 

\begin{figure}
    \centering
    \includegraphics[width=0.45\textwidth]{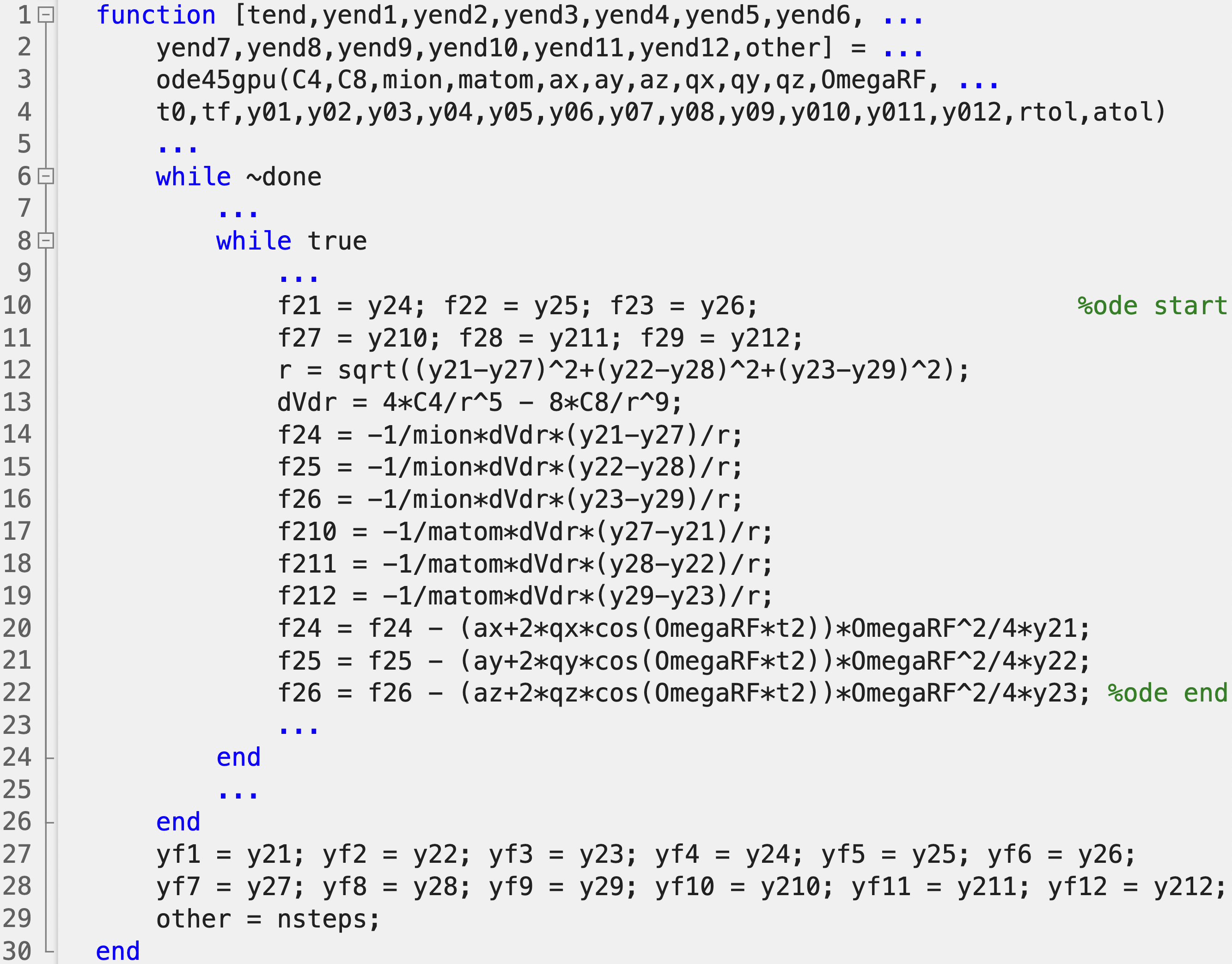}
    \caption{A snippet of code from ode45gpu showing an explicit expression for the force function $f(t,\bold{y})$. The lines of code corresponding to the force function, delineated by ``\%ode start'' and ``\%ode end'', are specific to the kind of interactions in the system. These lines of code appear seven times throughout ode45gpu, each with slightly different indices of the variables $f_i$. The ``...'' in lines 5, 7, 9, 23, and 25 indicate truncated lines of code.}
    \label{fig6}
\end{figure}

Given the definitions of each coordinate $y_i$ and the expression for $f(t,\bold{y})$, we wrote code for $f(t,\bold{y})$ without using any vectors, in a very particular way shown in Fig.~\ref{fig6}. The 12 components of $\bold{y}$ were written as \verb+y21+, \verb+y22+, ..., \verb+y212+. In the example shown in Fig.~\ref{fig6}, the 12 components of $f(t,\bold{y})$ were written as \verb+f21+, \verb+f22+, ..., \verb+f212+. Any kind of vector operation was written in terms of these variables; e.g., \verb+f(1:3) = y(4:6)+ was written as \verb+f21 = y24; f22 = y25; f23 = y26+. 

Next, we copied and pasted the code for $f(t,\bold{y})$ in seven places throughout the code for \verb+ode45gpu+. In each of the seven places, the \verb+f21+, \verb+f22+, ..., \verb+f212+ were replaced with, for example, \verb+f31+, \verb+f32+, ..., \verb+f312+.

Then, we specified the input and output arguments of \verb+ode45gpu+. As shown in line 1 of Fig.~\ref{fig6}, the outputs are the end time \verb+tend+, the final conditions \verb+yend1+, \verb+yend2+, ..., \verb+yend12+, and some other scalar observable \verb+other+. We later modified the output \verb+other+ to be some low-memory scalar observables, including the final kinetic energy of the ion, the lifetime of the complex, and the number of timesteps computed in the simulation. Each of these observables is calculated within \verb+ode45gpu+ using information from each timestep on the fly, or just the last timestep, without storing information from past timesteps. 

As shown in lines 2-3 of Fig.~\ref{fig6}, the input arguments to \verb+ode45gpu+ are the constants used to evaluate the force function (\verb+C4+ through \verb+OmegaRF+ in line 3 of Fig.~\ref{fig6}), the initial and final time of simulation \verb+t0+ and \verb+tf+, the initial conditions \verb+y01+, \verb+y02+, ..., \verb+y012+, and the relative and absolute tolerances \verb+rtol+ and \verb+atol+. 

\begin{figure}
    \centering
    \includegraphics[width=0.45\textwidth]{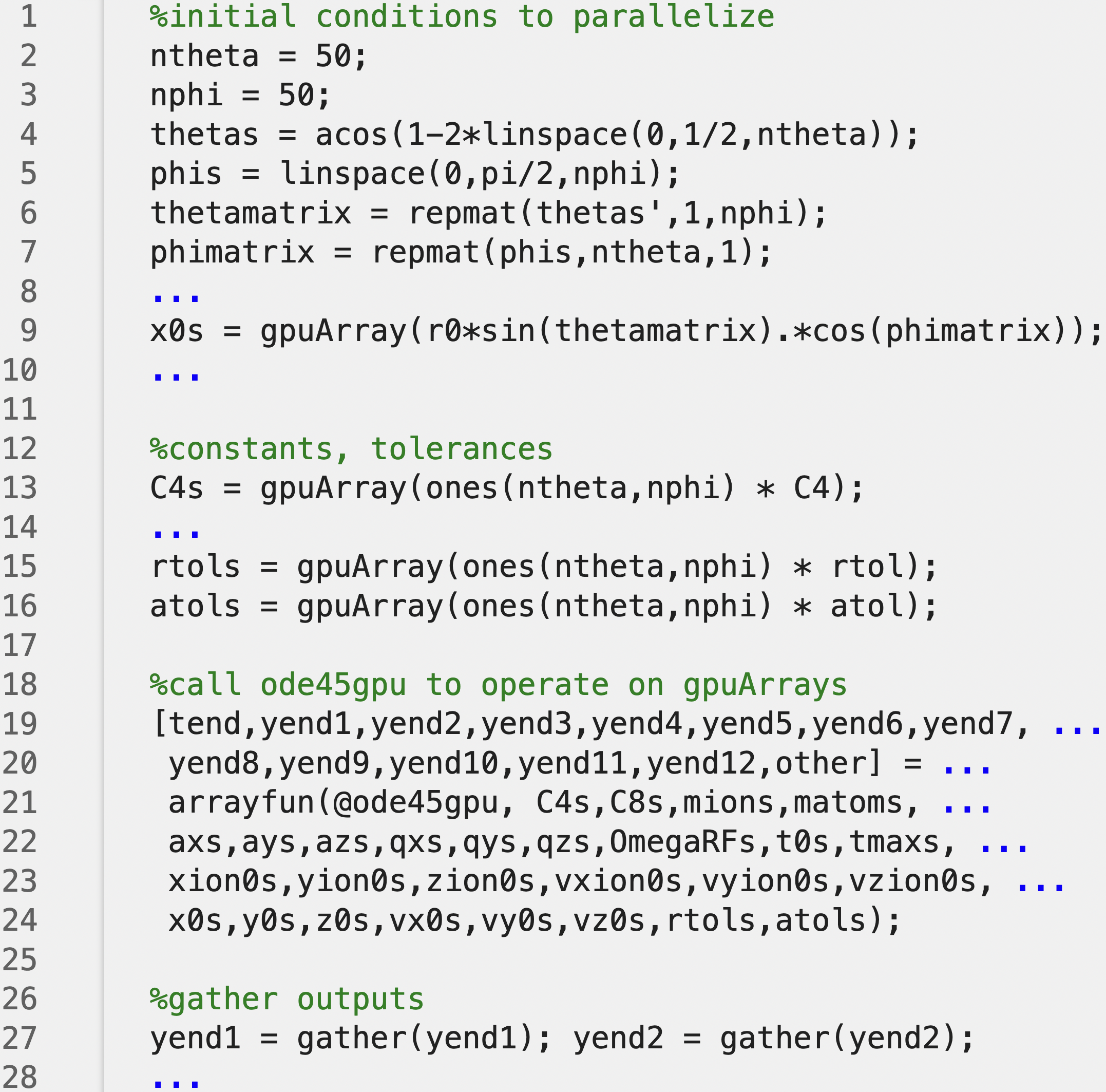}
    \caption{Example of how we use ode45gpu. First, for each initial condition to be parallelized, a matrix is constructed with that initial condition's values, and turned into a gpuArray. Then, for each constant parameter to be passed to ode45gpu, a matrix is constructed with all elements being equal to that constant. Next, arrayfun is used to call ode45gpu to operate on the gpuArray matrices on the GPU. Finally, gather is used to load the final results. Lines 8, 10, 14, 28 contain more lines of code hidden for brevity, while each ``...'' in lines 19-23 indicates that the code continues on the next line.} 
    \label{fig7}
\end{figure}

After setting up \verb+ode45gpu+, we supplied the initial conditions, as shown in Fig.~\ref{fig7}. We began by specifying matrices for the different initial spherical angles $\theta,\phi$, and then we wrote each initial condition (such as \verb+x0+, which corresponds to \verb+y07+) as a \verb+gpuArray+, as shown in line 9 of Fig.~\ref{fig7}. This was done for all 12 initial conditions; we are not showing the remaining 11 lines of code for brevity. Also, for each constant and tolerance, a \verb+gpuArray+ was constructed with the same dimensions as the initial conditions, where each \verb+gpuArray+ element was equal to that constant; see Fig.~\ref{fig7} lines 13-16. 

Then, we called \verb+ode45gpu+ using \verb+arrayfun+ as shown in Fig.~\ref{fig7} lines 19-24. This did an MD simulation on the GPU, numerically propagating $\bold{\dot{y}}=f(t,\bold{y})$, for each set of initial conditions given in the \verb+gpuArray+s for the inputs. 

Finally, as shown in Fig.~\ref{fig7} line 27, we used \verb+gather+ to gather each output of \verb+ode45gpu+, since the original outputs were also \verb+gpuArray+s (with the same dimensions as the initial condition \verb+gpuArray+s), but we prefer to analyze the data on the CPU. 

\subsection{Performance of ode45gpu on a GPU}

\begin{figure}
    \centering
    \includegraphics[width=0.45\textwidth]{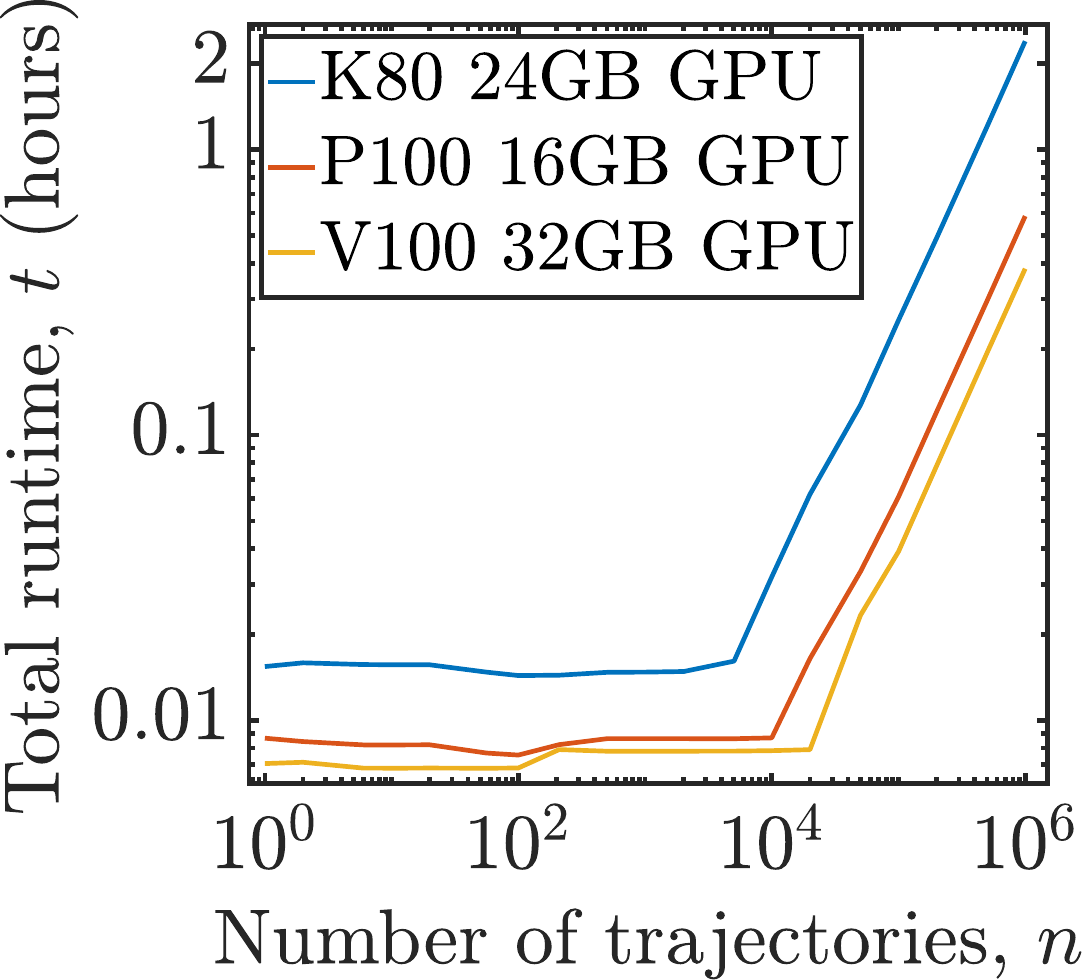}
    \caption{Runtime of ode45gpu on different GPUs, as a function of the number of trajectories $n\leq\,$1,000,000. For each $n$, the code ran $n$ identical copies of the specially-chosen trajectory shown in Fig.~\ref{fig1}(b).}
    \label{fig8}
\end{figure}

To benchmark the performance of \verb+ode45gpu+ on a GPU, we used three different NVIDIA Tesla GPUs: the K80 24GB, the P100 16GB, and the V100 32GB. For each GPU, we ran several different tests, where each test consisted of using \verb+ode45gpu+ on the GPU to run $n$ independent identical copies of the specially-chosen trajectory in Fig.~\ref{fig1}(b), where $n$ took on the values 1, 2, 5, 10, 20, ..., 1,000,000. The results of these tests are shown in Fig.~\ref{fig8}. For each GPU, there is a threshold number of trajectories near 10,000. Below the threshold, the code takes roughly the same amount of time to compute the trajectories regardless of $n$, which indicates that they are running in parallel. After the threshold, the runtime becomes roughly proportional to $n$, as expected with a limited amount of processing power. 

\begin{figure}
    \centering
    \includegraphics[width=0.45\textwidth]{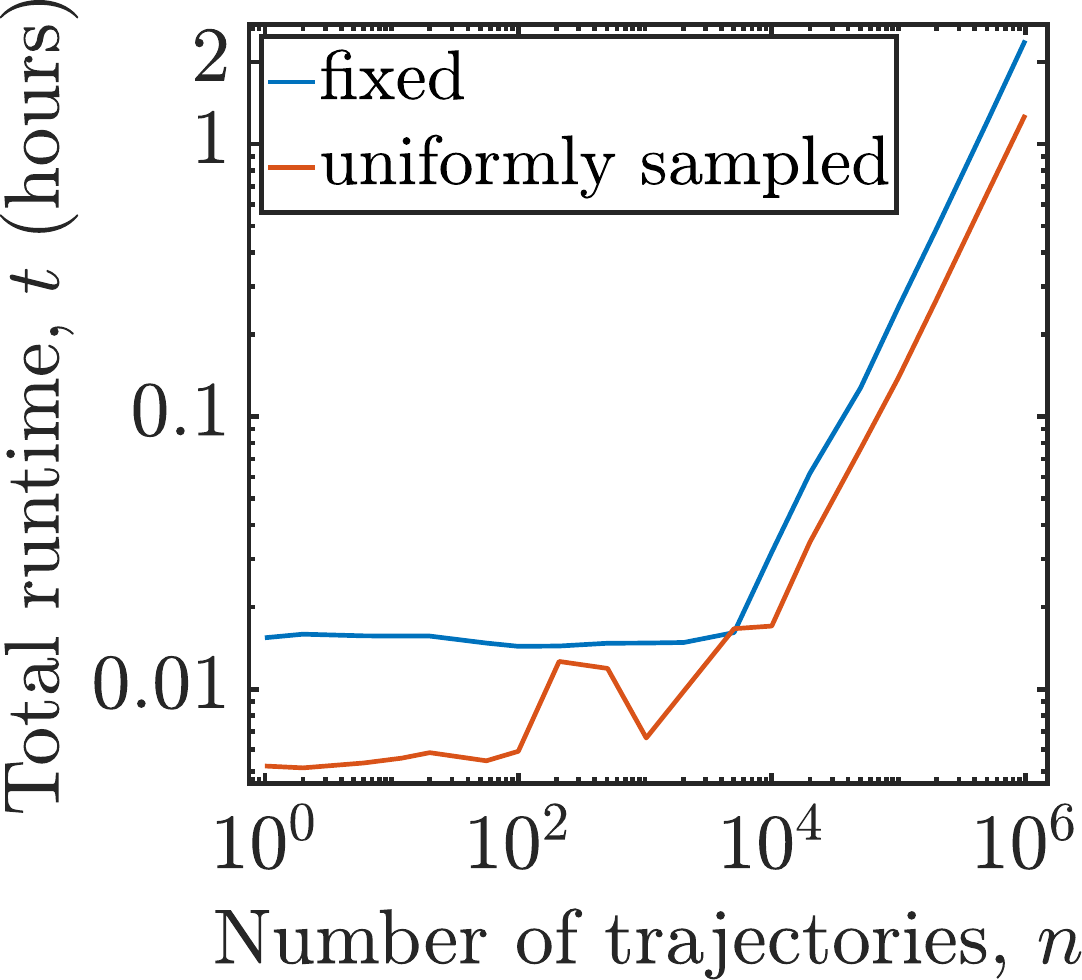}
    \caption{Runtime of ode45gpu on one Tesla K80 24GB GPU, as a function of the number of trajectories $n\leq\,$1,000,000. For each $n$, ``fixed'' represents running $n$ identical copies of the fixed, specially-chosen trajectory shown in Fig.~\ref{fig1}(b), and ``uniformly sampled'' represents running $n$ different initial conditions, sampled uniformly from the surface of the first octant of a sphere centered at the ion, $\theta,\phi\in[0,\pi/2]$.}
    \label{fig9}
\end{figure}

To be sure that the benchmark is not biased from all the trajectories being the same, we also tested \verb+ode45gpu+ with a more realistic scenario, running $n$ \textit{different} trajectories, with $n$ taking on the same values as before. More specifically, running only on the K80 GPU, we used \verb+ode45gpu+ to run different numbers of trajectories $n$, where for each $n$ the atom started at $n$ evenly-spaced positions (based on solid angle) on the first octant of the sphere of radius $r_0$ centered at the origin. The results of this test, compared to the results for the same GPU running $n$ identical trajectories, are shown in Fig.~\ref{fig9}. The uniformly-sampled case has more bumpiness in the runtime for small $n$, which makes sense because some of the trajectories might take much longer to simulate. After the threshold number of trajectories, both the fixed and uniformly-sampled cases show a runtime proportional to $n$, indicating that the differences between the trajectories average out.

\section{atomiongpu.m: a user interface for ode45gpu}

\begin{figure}[h]
    \centering
    \includegraphics[width=0.45\textwidth]{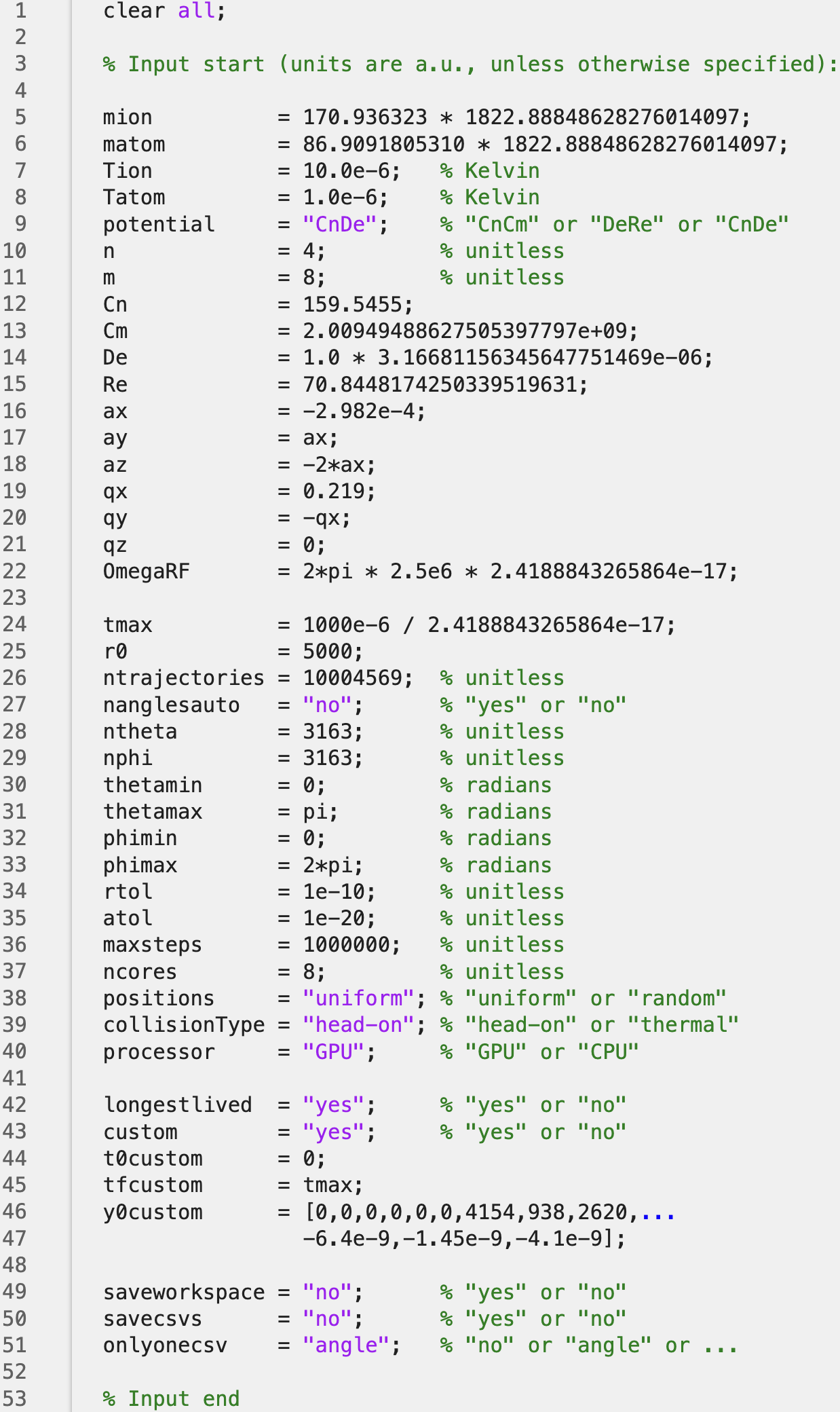}
    \caption{The complete set of inputs for the script atomiongpu.m. By default, the units are atomic units (a.u.), unless the comments say otherwise. For the string-based inputs, the comments show the different possible choices the user can use. The options for line 51 are given in the text. Some of the values given here are different from the code provided in the repository.}
    \label{fig10}
\end{figure}

In the previous subsection, we described how we used \verb+ode45gpu+ on the GPU from scratch. However, not everyone has the time to edit the code in detail, and we can automate the specific use case of a trapped ion interacting with a low-density cloud of atoms. So, to make \verb+ode45gpu+ easier to use, we developed a ``user interface'' script called \verb+atomiongpu.m+. The user only needs to change a few parameters appearing at the top, as shown in Fig.~\ref{fig10}, and the code does the rest. 

\subsection{Input to atomiongpu.m}

In Fig.~\ref{fig10} lines 5-6, the user specifies the masses in atomic units. The variable \verb+mion+ is the mass of the ion in atomic units, and is currently set to the mass of $^{171}$Yb$^+$. The variable \verb+matom+ is the mass of the atom in atomic units, and in this example it is the mass of $^{87}$Rb. Lines 7-8 give the temperatures of the ion and atom, in Kelvin. The precise meaning of these temperatures depends on whether \verb+collisiontype+ is set to \verb+"head-on"+ or \verb+"thermal"+. If \verb+collisiontype+ is set to \verb+"head-on"+, then the atom is launched towards the origin with speed $v_0=\sqrt{3k_BT_\textrm{atom}/m_\textrm{atom}}$, and the ion starts at rest at the origin (ignoring the variable \verb+Tion+). If \verb+collisiontype+ is set to \verb+"thermal"+, then each of the atom's velocity components is sampled from a normal distribution with standard deviation $\sqrt{k_BT_\textrm{atom}/m_\textrm{atom}}$, each of the ion's velocity components is sampled from a normal distribution with standard deviation $\sqrt{k_BT_\textrm{ion}/m_\textrm{ion}}$, and each of the ion's Cartesian coordinates is sampled from a normal distribution with standard deviation $\sqrt{k_BT_\textrm{ion}/m_\textrm{ion}/\omega^2}$, where $\omega$ is the trap's secular frequency in the $x$ direction, $\omega=\frac{1}{2}\Omega_\textrm{rf}\sqrt{a_x+\frac{1}{2}q_x^2}$. Basically, ``head-on'' means the atom is fired directly at the ion at fixed speed, while ``thermal'' means a more realistic scenario of an ion in a thermal bath of atoms.

Fig.~\ref{fig10}, lines 10-15 show how the atom-ion interaction is specified. A simple model for the interaction between an atom and ion is the Lennard-Jones type potential energy function $V(r)=-C_4/r^4+C_8/r^8$, where $C_4=\alpha/2$ and $\alpha$ is the polarizability of the atom, in atomic units, $C_8$ is another constant, and $r$ is the distance between the centers of mass of the atom and ion. A more legitimate model for $V(r)$ can be found, for example, by doing a quantum chemistry calculation at several different distances, and then running a cubic spline interpolation of the points. If the user wants to use their own model for the atom-ion interaction potential, they can first fit it to a generalized Lennard-Jones function of the form $V(r)=-C_n/r^n+C_m/r^m$, where $n$ and $m$ are positive integers and $C_n,C_m$ are positive constants expressed in atomic units. In this case, the user sets \verb+potential+ to \verb+"CnCm"+, and assigns values to \verb+n+, \verb+m+, \verb+Cn+, and \verb+Cm+ (the values of \verb+De+ and \verb+Re+ are ignored). If, instead, the user knows the potential depth $D_e$ and the long-range constant $C_n$, then the user sets \verb+potential+ to \verb+"CnDe"+, and assigns values to \verb+n+, \verb+m+, \verb+Cn+, and \verb+De+ in Hartree. The code then analytically calculates the value of \verb+Cm+ (ignoring the value given in line 13), and ignores \verb+Re+. Finally, if the user instead knows the potential depth $D_e$ and the equilibrium distance $R_e$, then the user sets \verb+potential+ to \verb+"DeRe"+, and assigns values to \verb+n+, \verb+m+, \verb+De+ in Hartree, and \verb+Re+ in Bohr radii. In this case, the code analytically calculates \verb+Cn+ and \verb+Cm+, ignoring the values given in lines 12-13. In every case, the atom-ion potential energy is $V(r)=-C_n/r^n+C_m/r^m$, where $n$ and $m$ are given in lines 10-11 and $C_n$ and $C_m$ are either calculated by the code or given in lines 12-13. 

\begin{figure*}[h]
    \centering
    \includegraphics[width=\textwidth]{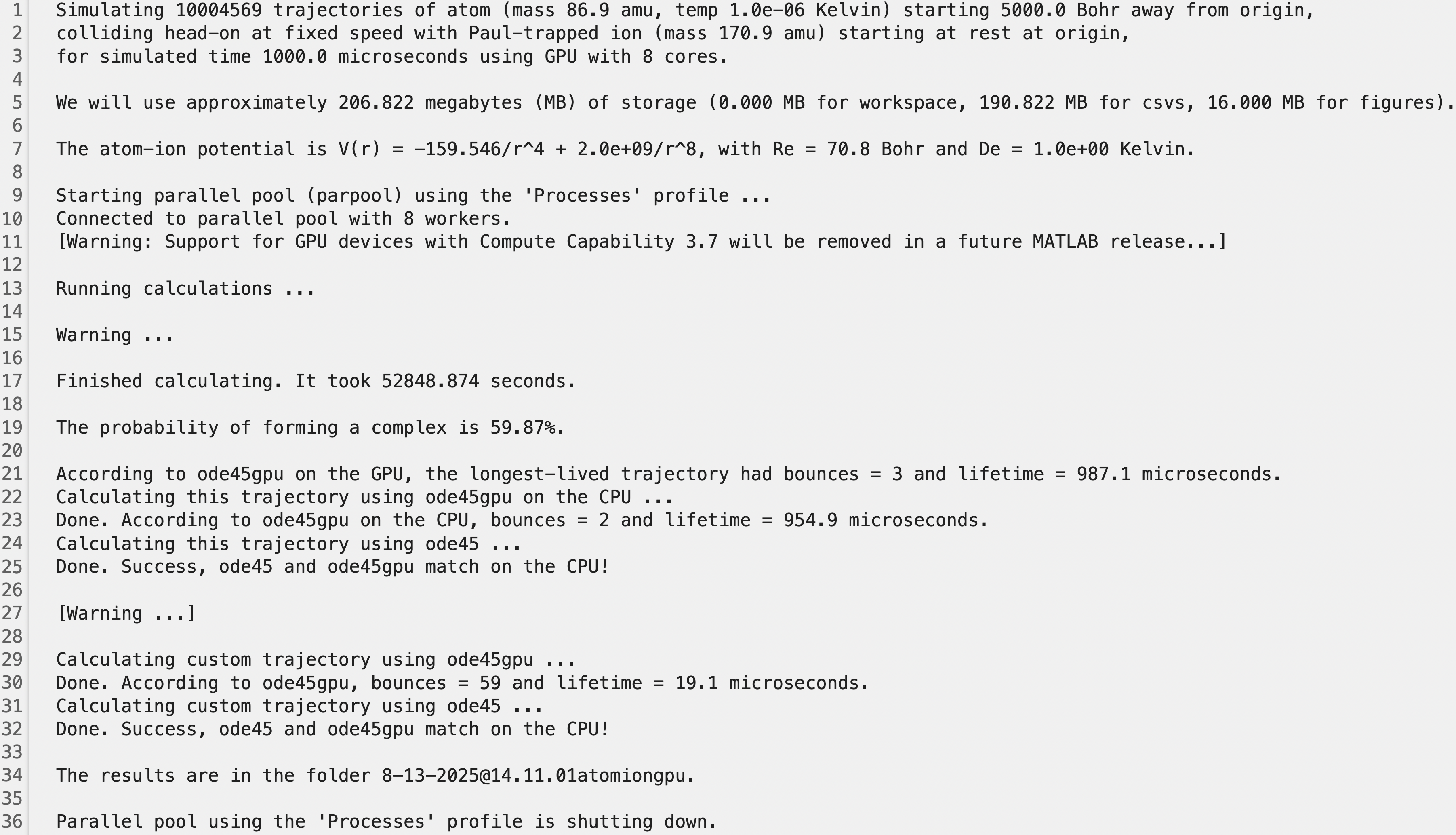}
    \caption{An example of the printouts when running atomiongpu.m. These are printed on the standard output, which is for example either the MATLAB command window, command line, or the usual output file if using a batch script. The warnings and parallel pool notifications are printed by MATLAB itself, while the remaining lines are printed by atomiongpu.m.}
    \label{fig11}
\end{figure*}

The code allows for the ion to be initially trapped in a Paul trap or harmonic trap, with trap parameters given in Fig.~\ref{fig10} lines 16-22. If the user does not want the ion to be trapped, all the variables \verb+ax+, \verb+ay+, \verb+az+, \verb+qx+, \verb+qy+, \verb+qz+, and \verb+OmegaRF+ should be assigned the value zero. If the user wants the ion to be trapped in a harmonic trap of the form $V_\textrm{trap}(x,y,z)=\frac{1}{2}m_\textrm{ion}(\omega_x^2x^2+\omega_y^2y^2+\omega_z^2z^2)$, then the user should assign \verb+qx=0+, \verb+qy=0+, \verb+qz=0+, \verb+OmegaRF=2+, \verb+ax+$=\omega_x^2$, \verb+ay+$=\omega_y^2$, and \verb+az+$=\omega_z^2$. Finally, if the ion is in a Paul trap, with potential $V_\textrm{trap}(x,y,z)=\sum_{i=x,y,z}(a_i+2q_i\cos\Omega_\textrm{rf}t)\frac{m_\textrm{ion}\Omega_\textrm{rf}^2}{8}r_i^2$, then the variables \verb+ax+, \verb+ay+, \verb+az+, \verb+qx+, \verb+qy+, \verb+qz+ (unitless), and \verb+OmegaRF+ (in atomic units) should be assigned the values of the corresponding Paul trap parameters.

Fig.~\ref{fig10} lines 24-40 are simulation and initial condition parameters. Each MD simulation terminates whenever the simulated time reaches \verb+tmax+ (in a.u., currently one millisecond), or whenever the number of timesteps reaches \verb+maxsteps+ (currently one million), whichever occurs first. The atoms are launched from a sphere of radius \verb+r0+ (currently 5,000 Bohr radii) centered at the origin. The number of different trajectories to be calculated is \verb+ntrajectories+, currently 10,004,569, and it can be set to any number up to about 10,000,000. Beyond that, there may be issues with memory, which we observed when trying to run 25,000,000 trajectories. The data for the trajectories is organized into \verb+ntheta+ $\times$ \verb+nphi+ matrices. If \verb+nanglesauto+ is set to \verb+"yes"+, then lines 28-29 are ignored, and the values of \verb+ntheta+ and \verb+nphi+ are picked automatically (each of them will be set to some number close to the $\sqrt{n_\textrm{trajectories}}$; the value of \verb+ntheta+ $\times$ \verb+nphi+ may exceed \verb+ntrajectories+ a little bit, so the code may simulate a few extra trajectories). Otherwise, if \verb+nanglesauto+ is set to \verb+"no"+, then the values of \verb+ntheta+ and \verb+nphi+ given in lines 28-29 are used, and in this case we require that \verb+ntheta+ $\times$ \verb+nphi+ $=$ \verb+ntrajectories+. 

The meaning of \verb+ntheta+ and \verb+nphi+ depends on the variable \verb+positions+ given in line 38. If \verb+positions+ is set to \verb+"uniform"+, then \verb+ntheta+ is the number of different values of spherical angle $\theta$, and \verb+nphi+ is the number of different values of spherical angle $\phi$ that are chosen for the initial position of the atom. \verb+nphi+ values of $\phi$ between \verb+phimin+ and \verb+phimax+ are chosen, evenly spaced. \verb+ntheta+ values of $\theta$ between \verb+thetamin+ and \verb+thetamax+ are chosen according to the formula $\theta=\arccos(1-2u)$, where $u$ takes on \verb+ntheta+ evenly spaced values between $(1-\cos\theta_\textrm{min})/2$ and $(1-\cos\theta_\textrm{max})/2$. Given the \verb+nphi+ values of $\phi$ and the \verb+ntheta+ values of $\theta$, all possible pairs of these values are used to construct \verb+nphi+ $\times$ \verb+ntheta+ ordered pairs of $(\theta,\phi)$ for the initial conditions of the atom. The cosines and arccosines are used to ensure that each resulting $(\theta,\phi)$ pair represents an equal area on the surface of the sphere. On the other hand, if \verb+positions+ is set to \verb+"random"+, then \verb+ntheta+ $\times$ \verb+nphi+ values of $\phi$ are chosen uniformly randomly between \verb+phimin+ and \verb+phimax+, and \verb+ntheta+ $\times$ \verb+nphi+ values of $\theta$ are chosen according to the formula $\theta=\arccos(1-2u)$, where $u$ is chosen uniformly randomly between $(1-\cos\theta_\textrm{min})/2$ and $(1-\cos\theta_\textrm{max})/2$. 

The variable \verb+ncores+, currently 8, specifies the number of CPU cores to parallelize over. The trajectories are distributed evenly across the \verb+ncores+ cores. If \verb+processor+ is set to \verb+"CPU"+, then each core runs its share of trajectories on the CPU, one trajectory at a time, using the \verb+ode45gpu+ function. Otherwise, if \verb+processor+ is set to \verb+"GPU"+, then each core will run its share of trajectories on a GPU, using MATLAB's \verb+arrayfun+ to parallelize \verb+ode45gpu+ with the different initial conditions on the GPU. In this case, ensure that \verb+ncores+ is less than or equal to the number of GPU's available on the machine, or else there may be an error.

\begin{figure*}[h]
    \centering
    \includegraphics[width=\textwidth]{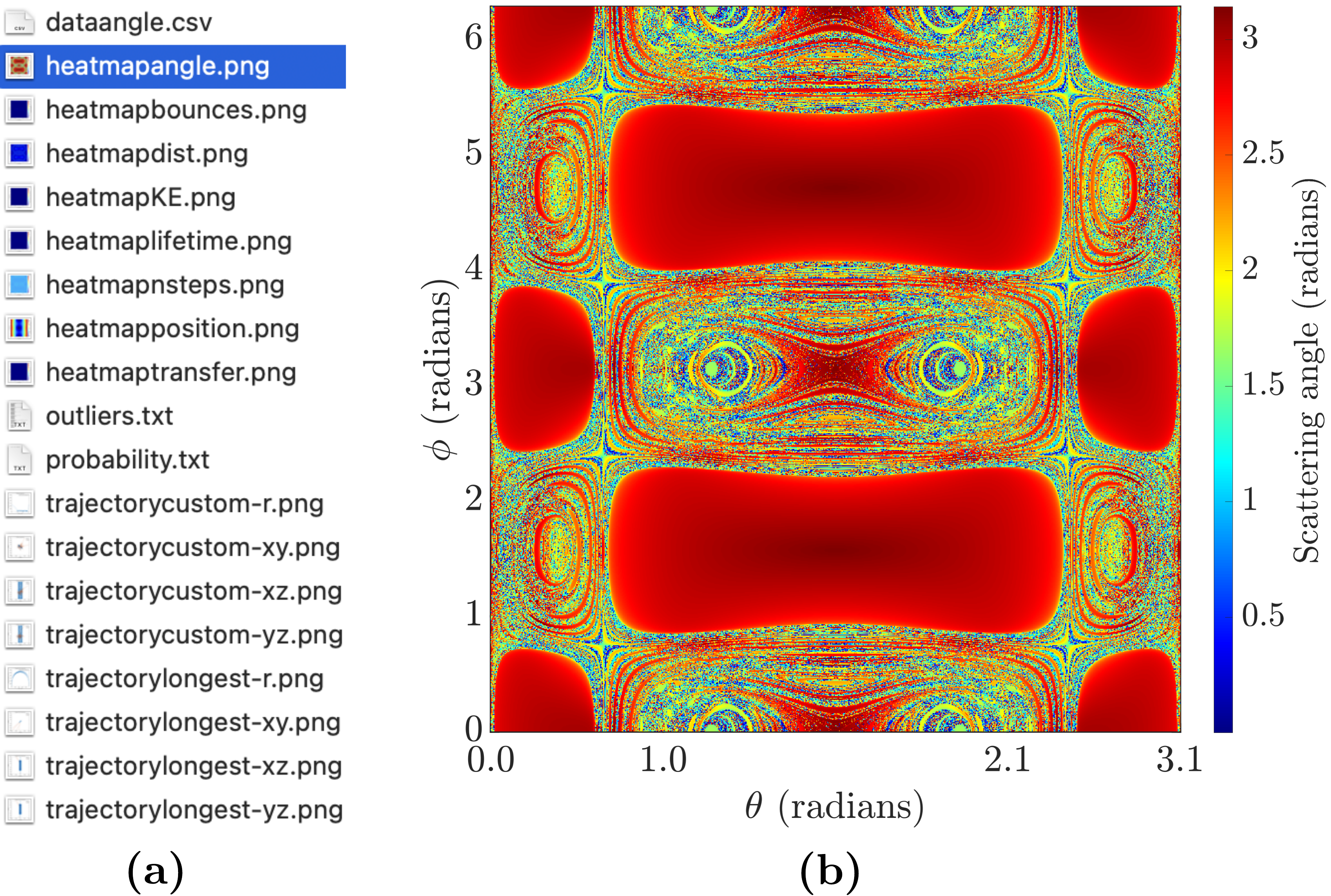}
    \caption{An example of the files generated as output by atomiongpu.m. atomiongpu.m creates a folder labeled by the current date and time, and within the folder, it stores several files, shown in Panel (a): .csv files of data, .png files of trajectories and heatmaps for observables, and .txt files containing other information. Panel (b) shows a heatmap of the scattering angle of the atom, as a function of the atom's initial position's spherical angles $\theta$ and $\phi$.}
    \label{fig12}
\end{figure*}

Fig.~\ref{fig10} lines 42-47 are parameters for displaying full trajectories. If \verb+longestlived+ is set to \verb+"yes"+, then the code finds which initial conditions had the longest complex lifetime, and again simulates the trajectory with the same exact initial conditions, using \verb+ode45+ this time. Then, the full trajectory is displayed in four different forms: a plot of the distance between the atom and ion as a function of time, an $xy$ projection of the trajectory (showing all the $(x,y)$ coordinates occupied by the atom and ion throughout the trajectory), an $xz$ projection, and a $yz$ projection. Otherwise, if \verb+longestlived+ is set to \verb+"no"+, this does not occur. Setting \verb+custom+ to \verb+"yes"+ does the exact same thing as setting \verb+longestlived+ to \verb+"yes"+, but with only one difference. Instead of using the initial conditions of the longest-lived trajectory, the custom initial conditions shown in lines 44-47 are used. In particular, the initial and final times are given by \verb+t0custom+ and \verb+tfcustom+, and the initial ion and atom positions and velocities are given by \verb+y0custom+, a 12-element vector where elements 1-3 are the ion's Cartesian coordinates, elements 4-6 are the ion's velocity components, 7-9 are the atom's Cartesian coordinates, and 10-12 are the atom's velocity components. The three dots \verb+...+ mean that the list of vector elements continues on the next line, line 47. 

Finally, for controlling the amount of memory used to save the final data, use Fig.~\ref{fig10} lines 49-51. To save the MATLAB workspace as a file \verb+work.mat+, set \verb+saveworkspace+ to \verb+"yes"+; otherwise, set it to \verb+"no"+. The workspace can later be opened to load all the parameters, initial conditions, and final conditions which were in memory when \verb+atomiongpu.m+ terminated. To save .csv files containing the raw data, set \verb+savecsvs+ to \verb+"yes"+ (in this case, the value of \verb+onlyonecsv+ is ignored); otherwise, set it to \verb+"no"+. To save only one observable's csv file of raw data, set \verb+savecsvs+ to \verb+"no"+ and set \verb+onlyonecsv+ to the name of the observable, which can be any one of these: \verb+"angle"+, which is the scattering angle of the atom; \verb+"bounces"+, which is the number of short-range collisions between the atom and ion throughout the trajectory; \verb+"lifetime"+, which is the lifetime of the atom-ion complex, or the amount of time between the first and last bounce; \verb+"position"+, which is how far the ion is from the origin during the first bounce; \verb+"transfer"+, which is the momentum transfered by the atom; \verb+"dist"+, which is the final atom-ion distance; \verb+"nsteps"+, which is the number of timesteps in the simulation; and \verb+"KE"+, which is the final kinetic energy of the ion. All observables are recorded in a.u. To save no csv files at all, set \verb+savecsvs+ to \verb+"no"+ and \verb+onlyonecsv+ to \verb+"no"+.

\begin{figure*}
    \centering
    \includegraphics[width=\textwidth]{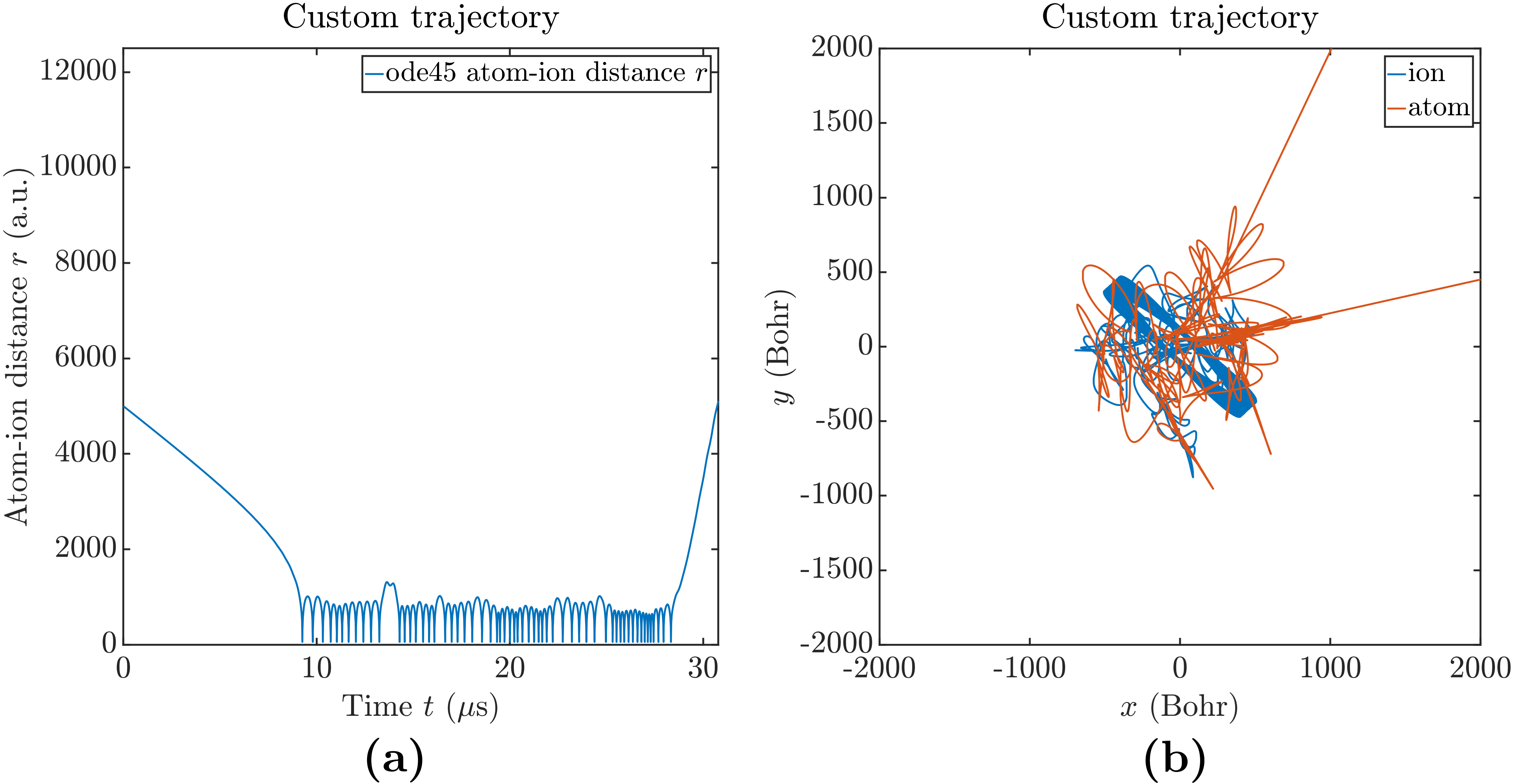}
    \caption{An example of a trajectory output. Panel (a) shows the atom-ion distance versus time for the first 30 $\mu$s of the custom trajectory. Panel (b) shows an $xy$ projection of the custom trajectory for the atom and ion.}
    \label{fig13}
\end{figure*}

\subsection{Output of atomiongpu.m}

After setting the input parameters, the code is ready to run. While it is running, it outputs a report of what is being calculated, shown in Fig.~\ref{fig11}. In lines 1-3, it gives a brief description of what is going to be simulated. In line 5, it estimates how much storage it will use to save the final data. If the user has less available space than the number stated in line 5, then the user should stop the program, set some of the variables in lines 49-51 in Fig.~\ref{fig10} to \verb+"no"+, and then start the program again. In line 7, the code states the atom-ion potential with rounded values of the coefficients. Lines 9-11 are automatically printed by MATLAB when connecting to the parallel pool of cores. Line 13 is displayed while the calculations are running, and when it finishes, it announces the elapsed time in line 17. In this case, using 8 GPU's, the code ran over 10 million trajectories in under 15 hours. In line 19, the code prints the percentage of trajectories which formed a complex. 

In Fig.~\ref{fig11}, lines 21-32 describe simulations of the longest-lived and custom trajectories, if they were requested in Fig.~\ref{fig10} lines 42-43. In this case, the code computes the longest-lived and custom trajectories, and reports its progress in Fig.~\ref{fig11} lines 21-32. It finds the simulated trajectory with the longest lifetime, and reports this trajectory's bounces and lifetime in line 21. If \verb+processor+ was set to \verb+"GPU"+ in Fig.~\ref{fig10} line 40, then the code recalculates the trajectory using the same function \verb+ode45gpu+ on the CPU, and reports its bounces and lifetime in Fig.~\ref{fig11} line 23. Note that lines 21 and 23 show different bounces and lifetimes, even though the initial conditions and ODE solver are exactly the same. This is because the CPU and GPU architectures are different, implemented with different round-off rules for floating point operations \cite{Whitehead2011}, leading to different results, especially when the dynamics is chaotic. As a way to double-check that \verb+ode45gpu+ is giving valid results, the code reruns the same trajectory using MATLAB's \verb+ode45+ on the CPU. If the final conditions, $y_1$ through $y_{12}$, match for \verb+ode45gpu+ and \verb+ode45+, then the code writes a success message on line 25; otherwise, it writes a failure message. The custom trajectory is handled in a similar manner. The code first calculates this trajectory on the CPU with \verb+ode45gpu+, reporting the bounces and lifetime in line 30, and then double-checks by recalculating it with \verb+ode45+ on the CPU. 

The warnings in lines 11, 15, and 27, abbreviated for brevity, can be ignored. The warning in line 11 means that in a future version of MATLAB, the code might not work with older GPUs. We used MATLAB version 2023a. 

The code not only prints information to the standard output, but it also creates figures and saves them in a folder. As shown in Fig.~\ref{fig11} line 34, the name of the folder contains the current date and time. An example of the files that the code stores in the folder are shown in Fig.~\ref{fig12}(a). Since \verb+savecsvs+ is set to \verb+"no"+ and \verb+onlyonecsv+ is set to \verb+"angle"+, Fig.~\ref{fig12} shows only one csv file, ``dataangle.csv''. 

For each observable, the code also converts the raw data to a heatmap, showing that observable as a function of the spherical angles $\theta$ and $\phi$. For example, Fig.~\ref{fig12}(b) shows the image ``heatmapangle.png'', which is a heatmap for the atom's scattering angle. This image shows that there are certain regions of incoming angles where the atom simply back-scatters, and there are other regions with more chaotic dynamics, producing a fractal structure. For the custom and longest-lived trajectories, the code produces figures depicting the actual trajectories. For example, the files ``trajectorycustom-r.png'' and ``trajectorycustom-xy.png'' show an atom-ion-distance vs. time plot and an $xy$ projection of the trajectory, as shown in Fig.~\ref{fig13}. 

The file ``probability.txt'' just stores the probability of complex formation, and the file ``outliers.txt'' stores the initial conditions for the trajectories with the highest and lowest value of each observable. These initial conditions can be copied and pasted into lines 46-47 of Fig.~\ref{fig10} as custom trajectories to make the code show what they look like. 

\section{Conclusion}

In this paper, we presented our MATLAB function \verb+ode45gpu+, which propgates ordinary differential equations describing atom-trapped-ion dynamics, on either a CPU or GPU, and runs 22x faster than MATLAB's \verb+ode45+. We then provided a script called \verb+atomiongpu.m+, which makes it easy for the user to customize all the parameters for the atom, ion, trap, ODE propagation, initial conditions, and parallelization, letting the code do the rest, taking advantage of the available hardware to simulate up to about 10 million trajectories at a time, and automatically outputting .csv files and heatmaps of the final conditions, trajectory plots, and other information like the probability of complex formation. Those working in the field of ion trapping and ion-neutral interactions will find this script to be a convenient tool for quickly testing different trap parameters before trying them in the laboratory, understanding their experimental results afterwards, and exploring ion heating and chaotic atom-trapped-ion dynamics.

\section{Acknowledgements}

This work was supported by the United States Air Force Office of Scientific Research [grant number FA9550-23-1-0202].

\bibliographystyle{plainnat}

\end{document}